# PROTECTING PRIVACY IN VANETs USING MIX ZONES WITH VIRTUAL PSEUDONYM CHANGE

"The author declares that there is no conflict of interest regarding the publication of this paper."

Belal Amro

Hebron University

## ABSTRACT

*Vehicular ad hoc networks VANETs use pseudonyms to communicate among them and with roadside units, these pseudonyms are used to authenticate these vehicles and to hide real identities behind these pseudonyms, to better enhance privacy, these pseudonyms are changed frequently so that it will not be that easy to link these pseudonyms together and hence reveal real identities. However, changing pseudonyms will not be that useful if previous and current pseudonyms are easily linked together. Therefore different techniques have been proposed to hide the pseudonym changes and make it difficult to link pseudonyms together. Most of these techniques do not fully quarantine privacy when changing a pseudonym under some situations such as low traffic. In this paper, we provide a technique for changing pseudonyms that have the same privacy level under all traffic conditions. The technique relies on fixed mixing zones that are planted and distributed over the roads. By this technique, a vehicle guarantees a high level of security when changing its pseudonym at that mix zone which will make it very difficult for an adversary to link particular pseudonyms together and hence reveal real identity. Performance analysis showed that our model works efficiently with very few computational costs.*

## KEYWORDS

*VANETS, privacy, pseudonyms, mix zone, security.*

## 1. INTRODUCTION

The Vehicular Ad-Hoc Network, VANET, is a technology that uses moving vehicles as nodes in a network to create a mobile network. Each vehicle takes on the role of a sender, receiver, and router to broadcast information to the network. This information is then used to ensure the safe and free flow of traffic. Vehicles are equipped with some sort of radio interface called On Board Unit (OBU) that enables communication ; with other vehicles and with Road Side Units (RSU). Vehicles are also equipped with hardware that permits detailed position information such as Global Positioning System (GPS). Fixed RSUs, which are connected to the backbone network, must be in place to facilitate communication, this model is shown in Figure 1.

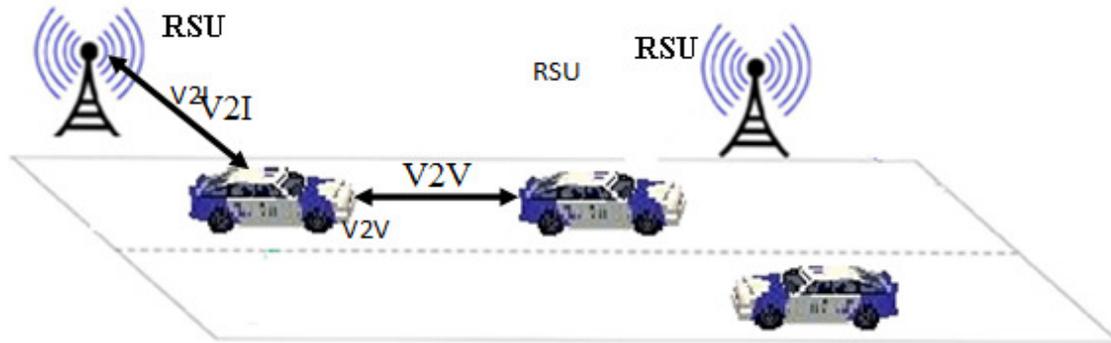

Figure 1: VANETs architecture

One of the mechanisms used for Authentication in VANETS is a digital signature. Because of a large number of network members and variable connectivity to authentication servers, a Public Key Infrastructure (PKI) is used for authentication where each vehicle would be provided with a public/private key pair. These Public keys are called pseudonyms [4,5].

To enhance privacy in CTM systems, users use temporary identities called pseudonyms instead of using their real identities [6]. The use of pseudonyms enables hiding the real identity of users. However, it has been showed that even when using pseudonyms, it may be possible to reveal the real identity of some users by their location information [17].

CTM system requires users to update their location information at the traffic server from time to time so that the traffic server will be aware of traffic conditions. Different update mechanisms have been proposed in the literature. These update mechanisms vary according to their update frequencies and time gaps between successive updates. The first approach is periodic updates where location information is updated periodically at fixed time intervals [13], the other technique is a conditional update where location update occurs only if a vehicle crosses a boundary [3,9].

To overcome this flaw, users are asked to change their pseudonyms from time to time. Changing pseudonyms makes it difficult to link pseudonyms together with the aim to build partial or even complete trajectory of the moving object. This in turn enhances the privacy level of users. However, pseudonym change is of no use if very few users do this change. Few users means higher probability of linking corresponding pseudonyms successfully[5,7].

In this paper, we propose a technique of securely changing pseudonyms that fully guarantees high privacy level for the vehicles that change their pseudonyms in a particular region called mix zone[5,7]. The privacy level remains high and is not affected by number of users inside the mixing zone.

The rest of this paper is organized as follows; in Section 2 we have provided a detailed study of related work. In Section 3 we have explained our proposed method. Discussion and Performance evaluation is provided in Section 4. At the end, Conclusions and Future works are provided in Section 5.

## 2. RELATED WORK

A secure dedicated infrastructure (DI) system architecture based on the use of pseudonyms for identity hiding has been proposed by Raya et al. [15]. Pseudonyms are temporary identifiers that expire after their use. Many variations of pseudonyms usage in VANETs were proposed [23, 24]. Despite the fact that pseudonyms are temporary identities, it has been shown that it is possible to track pseudonyms changes and disclose real identities by using some probabilistic models as described in[5].

One major solution for linking pseudonyms was using mix zones. The concept of mix zones has been proposed by [14] where they defined mix zones as hidden areas where users can change their pseudonyms together without being linked. In this way, the pseudonym change is hidden, and hence pseudonyms will not be linked easily. Pseudonyms and mix zones have been proposed to be used in Dedicated Infrastructures (VANETs). However, DI approach is still under research [1,4,13, 25] and no important real deployments exist.

Another major solution for protecting moving objects privacy is called online privacy preserving, Many approaches have been proposed for protecting the privacy of location data. One technique is called *cloaking*, either time, space, or both [3,10,12]. Cloaking is a process of generalization, where time or space is expanded so that a $k$-anonymity level is met, $k$-anonymity refers to the state of being anonymous among other $k$ objects [16]. Cloaking gains a guaranteed $k$-anonymity level on the account of the quality of the data. As a drawback, all cloaking techniques rely on either a trusted third party that determines the boundaries of the cloaked region [11], or on collaboration among users to update their locations together [12]. The latter relies on direct communication between group members, and requires trust between them.

Hoh and Gruteser provided the concept of Virtual Trip Lines (VTL) [3]. A VTL is a geographic boundary that is supplied to the client software, and a vehicle must update her location upon the cross of that boundary. The system fails to capture traffic conditions apart from VTL regions, because vehicles will not update their locations outside VTL regions. Besides, if some VTL lines are compromised then it will be easy to link pseudonyms at these compromised VTLs. The very much effort done on the choice of VTLs and their distribution to users makes the system impractical for a larger number of users.

In [11], the authors use an anonymization server that anonymizes user locations using location cloaking, where location information is perturbed by either spatial range called spatial cloaking, or temporal range called temporal cloaking. Thus, exact location information is hidden among a range of temporal and spatial coordinates. It guarantees $k$-anonymity in both time and space dimensions. But it still relies on a trusted third party and also degrades the quality of the data. In [8], the authors suggest a 2 way cloaking mechanism, the user sends her cloaked location to an anonymization server, the latter returns a cloaking rectangle that has $k$ other users. Cloaking relies on a trusted third party for calculating the safe region; it also reduces the data quality by generalizing location into the safe region. A similar approach was used by [12] where users communicate together to form a safe region without communicating with a trusted third party, in the latter approach users are assumed to be honest and trust each other to calculate the safe region.

As a general description of online privacy-preserving, all techniques require trusted third party and extra communication to establish the mixing zone or to cloak the data [26]. Amroet. Al. [2], have proposed a technique that automatically and autonomously creates a mixing zone without the need of a trusted third party. However, this method is still prone to privacy leakage when under low traffic conditions.

**Proposed model:**

In this section we will explain our proposed high privacy technique for changing pseudonyms in a mix zone, first we will show the structure of our mix zone, and then we will explain how this zone will be operated to leverage privacy level. At the end we will provide some suggestions that are necessary to guarantee the privacy of users according to k-anonymity metric.

**Uses of mix zones:**

Mix zones for a group of users are defined by [18] as a connected spatial region of maximum size in which none of these users has registered any application callback. This definition might be extended for VANETs as a region where vehicles can change their pseudonyms safely where it becomes impossible for intruders to track this change in sake of real identity reveal.

To maximize the efficiency of a mix zone, it is highly recommended that vehicles change pseudonyms in mix zones where sufficient number of vehicles do that change to make it difficult to link previous pseudonyms of different vehicles with new ones. Therefore implementing real mix zone will be efficient in places where we have a sufficient number of vehicles such as traffic lights and squares.

However, it is not always guaranteed to have the required number of vehicles to change these pseudonyms together in a mix zone, and hence privacy might be invaded. Therefore, to serve users of high privacy needs we propose a model that works efficiently though a sufficient number of vehicles was not met.

**Mix zone with virtual changes**

We used the model of [19] where vehicles change their pseudonyms in a predefined regions, Figure 2 shows a graphical representation of a mix zone. The added value of our model is that a vehicle can safely change pseudonym in lack of sufficient number of neighboring vehicles. We introduce the concept called virtual change where pseudonym change is performed by non-real vehicles, and hence the adversary will consider it in his calculation to link pseudonyms.

Figure 2 shows the structure of the mix zone, where we see the gates which denote the entry/exit point of the mix zone. The term Ingress refers to vehicles entering mix zone from a particular gate, while the term Egress refers to one leaving the zone.

International Journal of Network Security & Its Applications (IJNSA) Vol. 10, No.1, January 2018

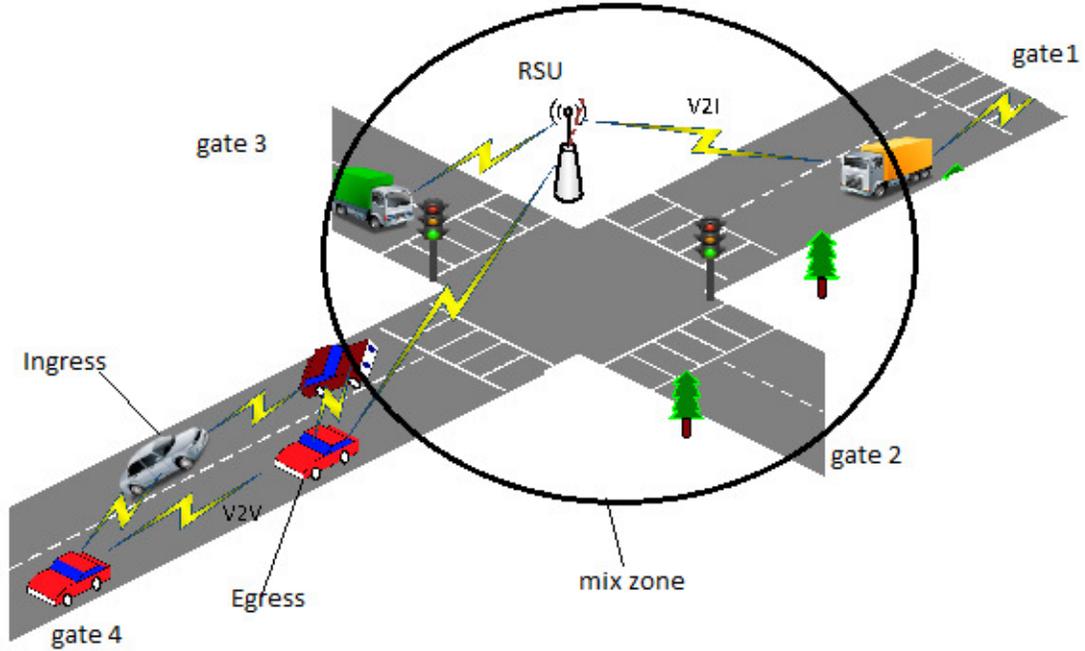

Figure 2: Mix zone structure

Our enhancement for the model in Figure 2 is the deployment of the gates. At each gate we deployed 20 transceivers, 10 for each direction. Each transceiver is 10 meters away from the other. And these transceivers will be called virtual vehicles and their role is to mimic the process of changing pseudonyms. Each transceiver will communicate with RSU like a real vehicle with a particular pseudonym. By making these transceivers do that job on all gates and all directions then we will have a complex model for an adversary to analyze. Figure 3 shows the structure of each gate.

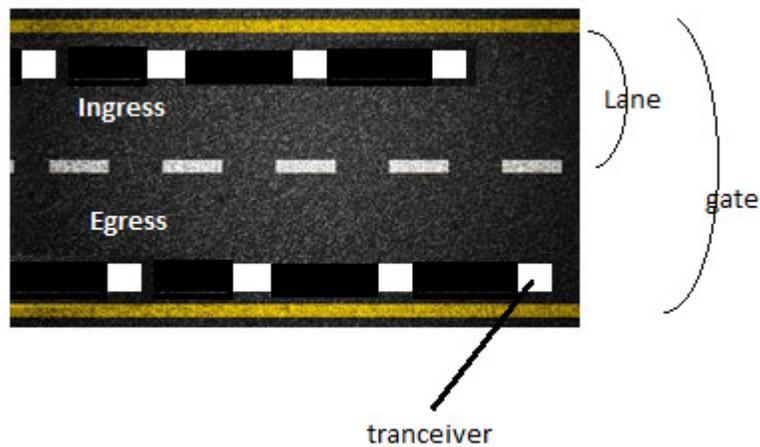

Figure 3: composition of a gate in the mix zone

According to Figure 3, we will be having a set of transceivers at both directions for each gate, these transceiver will be working together and will perform pseudonym change process that includes:

1. Communication using an old unused pseudonym before mix zone entrance

2. Encrypted communication inside mix zone

3. Communication using a new pseudonym at the most probable exit gate.

Performing these steps will make it very difficult for an adversary to link old and new pseudonyms together by mixing real vehicle changes with these virtual changes. Further details about the process are discussed in Performance analysis section.

**Proposed algorithm**

According to [20], an adversary will be powered by a movement matrix that is used to calculate the probability of exit from a particular gate according to the entry gate and time stamp. By this matrix the probabilities are calculated, These probabilities help adversary to link old pseudonym with new one.

From this probability matrix, a mapping is calculated and probability is associated with each mapping. If we have enough vehicles changing their pseudonyms inside the mix zone, then we will be having a sufficient number of mappings with high probability that it will make it difficult for adversary to link pseudonyms.

To proceed with our method we need to keep track of current state of the mix zone, i.e. number of vehicles at all gates of the mix zone. This information will be used to calculate the weights of different mapping possibilities between gates of the mix zone.

Our proposed method is described in the following algorithm:

*Input: state_matrix, EG_ING_prob_Matrix*

*Calculate W_MAP matrix*

*For all W_MAP matrix entries*

*If(W_MAP[$gate_i$][$gate_j$]<W_MAP$_{Thresh}$)*

*Then*

    *Activate (10- state_matrix[$g_i$][$l_i$]) transceivers at $gate_i$ ingress lane*

    *Activate (10- state_matrix[$g_i$][$l_i$]) transceivers at $gate_j$ egress lane*

Where :

- *state_matrix: is a matrix containing number of vehicles entering and leaving miz zone at each lane. i.e. state_matrix[2,1]=4 means that we have 4 vehicles leaving gate 2 while state_matrix[2,2]=4means that we have 4 vehicles entering mix zone at gate 2.*

- *EG_ING_prob_Matrix: is a matrix that contains the probabilities of linking gate lane together. i.e. if EG_ING_prob_Matrix[2,3]=0.2 means that the probability of a vehicle entering at gate 2 will leave at gate 3 is 0.2.*

- *W_MAP: is a matrix that contains the weights of mapping between gates, this value is affected by both number of vehicles entering in a particular gate and the number of vehicle leaving at the other gate besides being affected by the probability of linking the two gates. i.e. W_MAP[1][3] means the weight of mapping pseudonym entering from gate one to a pseudonym leaving at gate 3 and is calculates as : $W\_MAP[1][3] = (state\_matrix[1][2]^{state\_matrix[3][1]}) * EG\_ING\_prob\_Matrix[1,3]$. That is to say the weight of a mapping is calculated as number of entering vehicles at a gate raised to the power of number of vehicles leaving at the other gate then the value is multiplied by the probability that a vehicle will leave through that gate.* For better analysis we used the normalized value where we divided the previous calculated value by the sum of the calculated value for each egress.

Note that we will run this algorithm once we have less than 10 vehicles at any lane of each gate. Once we have 10 or more vehicles for that lane we will not activate transceivers over that lane. It is worth mentioning here that the probability of linking gate lanes is also considered in calculating

## 3. DISCUSSION PERFORMANCE ANALYSIS

Vehicles entering mix zone must change their pseudonyms inside the mix zone and leave the mix zone with a new pseudonym. The power of the adversary depends on how successful can he map an old pseudonym with the new one of the same vehicle.

As the number of vehicles entering and leaving the mix zone from different gates at suitable amounts of time periods increase the difficulty of mapping two successive pseudonyms of a particular vehicle decreases. Assuming that at time $t_0$ we have 10 vehicles entering the mix zone at gate 1 and at time $t_1$ we have 10 vehicles leaving the mix zone from the 3 other gates in Figure 2. Then we have at most 3*10! Possible mappings.

Some of these mapping might be ruled out because of many reasons including:

1. Vehicles wait more time in presence of traffic lights

2. Some gates might be excluded due to time constraint (i.e. it is impossible for a vehicle entering at gate 1 to exit at gate 4 in $t_1$-$t_0$ time interval

3. The mix zone property might yield to ignore some exits depending on some historical data (if gate 3 exit occurs rarely from vehicles entering at gate 1, then these mapping might be excluded).

Despite all the above mentioned facts, we still have a very powerful indication of privacy level depending on the values obtained by *W_MAP* matrix. The calculation of *W_MAP* is straight forward and does not require that much complexity, it depends on the number of gates of the mix zone which is limited and might not exceed 8 gates in most cases. So we are having a mix zone of n gates then calculating W_MAP will require $O(n^2)$. Activating transceivers will take a constant time and hence the total algorithm will run in $O(n^2)$ where n is the number of gates which is limited by 8 in most cases.

As an illustrating example if we have the state matrix and *EG_ING_prob_Matrix,* shown in Table 1 and Table 2 respectively, .the calculations will proceed as follows*:*

Table 1: state matrix

|         | Gate1 | Gate2 | Gate3 | Gate4 |
|---------|-------|-------|-------|-------|
| Ingress | 10    | 3     | 6     | 8     |
| Egress  | 7     | 10    | 9     | 8     |

The entries of the state matrix represent the number of real vehicle at each lane of the gate, for example *state_matrix[1][3]=6* means that we have 6 vehicles entering at gate 3.

Table 2: *EG_ING_prob_matrix*

|       | Gate1 | Gate2 | Gate3 | Gate4 |
|-------|-------|-------|-------|-------|
| Gate1 | 0.01  | 0.30  | 0.30  | 0.39  |
| Gate2 | 0.19  | 0.01  | 0.40  | 0.40  |
| Gate3 | 0.39  | 0.10  | 0.01  | 0.50  |
| Gate4 | 0.60  | 0.09  | 0.30  | 0.01  |

The entries of this matrix represents the probability of linking two lanes of different gates, for example *EG_ING_prob_matrix[2][3]=0.4* means that a vehicle entering at gate 2 will leave at gate 3 with probability 0.4.

Table 3: *W_MAP* matrix

|       | Gate1   | Gate2 | Gate3 | Gate4 |
|-------|---------|-------|-------|-------|
| Gate1 | 0.00003 | 0.898 | 0.089 | 0.117 |
| Gate2 | 0.036   | 0.051 | 0.684 | 0.228 |
| Gate3 | 0.015   | 0.852 | 0.014 | 0.118 |
| Gate4 | 0.009   | 0.699 | 0.291 | 0.001 |

The entries of the W_MAP matrix represent the weights of different mapping among mix zone lanes. The higher the value the better privacy and the more difficult for adversary to link pseudonyms between these lanes.

As seen from Table 3, it is clearly obvious that the W_MAP[1][3] is low compared to the high probability associated with it (0.3) hence it might be better to activate transceivers on egress lane of gate 3. Utilizing W_MAP matrix will enable us to pinpoint the privacy invasion scenarios and will be able to act accordingly to enhance privacy and make it harder for an adversary to link pseudonyms.

**Optimization techniques**

To achieve better results, we recommend the following :

The use of the same transceiver types of the same company to reduce the effect of radio frequency.

finger prints. You may refer to [21] for more details about radio frequency fingerprint identification.

Design the mix zone in a way that we get the same probabilities for all possible linking, i.e. identical value for all entries in the *EG_ING_prob_matrix*. This might be achieved by careful design of the mix zone with same distances among all gates. For more details about optimal placement of mix zone you may refer to [22].

## 4. CONCLUSIONS AND FUTURE WORK

Privacy preserving techniques using changing pseudonyms may lead to privacy leakage by having the possibility of linking previous and current pseudonyms and hence reveal real identities. Mix zones have been proposed to enable users to change their pseudonyms secretly, however it has been proven that under some situation this pseudonym change might be tracked by statistical models. In this paper, we proposed a mix zone model that is aware of these situations and interact positively to enhance privacy. Performance analysis showed that our model preserves a privacy level with constant time complexity.

As a future work, we intend to generalize our model for all location based service application other than VANETs, besides we intend also to make that model more and more flexible and suitable for these applications.

.